\documentclass[12pt,preprint]{aastex}

\slugcomment{Astrophysical Journal, in press, 10 July 2006, v645n 2 issue.}

\shorttitle{Phase in Nulling Interferometry}
\shortauthors{Danchi, Rajagopal, et al.}

\begin{document}

\title{The Importance of Phase in Nulling Interferometry  \
and a Three Telescope Closure-Phase Nulling Interferometer Concept}
\author{W. C. Danchi, J. Rajagopal\altaffilmark{1}, M. Kuchner, J. Richardson, D. Deming\altaffilmark{2}}
\affil{NASA Goddard Space Flight Center, Exoplanets and 
Stellar Astrophysics Laboratory, Code 667, Greenbelt, MD 20771}
\email{William.C.Danchi@nasa.gov}
%
\altaffiltext{1}{University of Maryland, Astronomy Department, College Park,
MD 20742}
\altaffiltext{2}{Planetary Systems Laboratory, Code 693, Goddard
Space Flight Center, Greenbelt, MD 20771}
\begin{abstract}
We discuss the theory of the Bracewell nulling interferometer and 
explicitly demonstrate that
the phase of the ``white light" null fringe is the same as the 
phase of the bright output from an ordinary stellar interferometer.
As a consequence a ``closure phase" exists for a nulling interferometer
with three or more telescopes. 
We calculate the phase offset as a function of baseline length for an
Earth-like planet around the Sun at 10 pc, with a contrast ratio
of $10^{-6}$ at 10 $\mu$m.  The magnitude of the phase due to the
planet is $\sim 10^{-6}$ radians, assuming the star is at the phase
center of the array.  Although this is small, this phase may
be observable in a three-telescope nulling interferometer 
that measures the closure phase.
We propose a simple non-redundant
three-telescope nulling interferometer that can perform
this measurement.  This 
configuration is expected to have improved characteristics compared
to other nulling interferometer concepts,
such as  a relaxation of pathlength tolerances,
through the use of the ``ratio of wavelengths" technique, a 
closure phase, and better discrimination between exodiacal
dust and planets.

\end{abstract}

\keywords{telescopes --- techniques: interferometric ---
techniques: high angular resolution ---
stars: planetary systems --- stars: circumstellar material}

\section{Introduction}
Direct imaging
of Earth-like planets around nearby stars is extremely difficult
for two fundamental reasons. The first is the need for 
resolution well below one arcsec; the second is the extremely large
contrast ratio between the planet and star, which is  
$\sim 10^{-6} - 10^{-7}$ at mid-infrared wavelengths (e.g., 5-20 $\mu$m)
and  $\sim 10^{-9} - 10^{-10}$ at visible wavelengths (400-900 nm).  Generally, 
interferometric techniques are favored in the infrared, while coronagraphic techniques have received the most attention at visible wavelengths.
The extrasolar zodiacal dust around these stars provides an additional
complication, primarily 
because their intensity compared to the zodiacal dust in our
own Solar system, and also their spatial distribution, is largely unknown.

Interferometric methods for direct detection of extrasolar planets are
derived from the initial concept of Bracewell (1978) (see also 
Bracewell \& MacPhee 1979) of a rotating
two-telescope interferometer, in which a 180 degree phase shift was 
applied to the electric field from one of the two telescopes.  The 
net result of this phase shift is a response pattern with a minimum
or ``null" response on-axis (zero pathlength difference
between the two telescopes),
which suppresses the unwanted signal from
the starlight, and has a maximum response off-axis, at an angle 
proportional to the wavelength of light divided by the separation
between the two telescopes.  

More than a quarter century has passed since that initial paper, and
there has been a substantial body of work in which a variety of 
array configurations of interferometers has been proposed for planet detection, 
including the OASES (Angel \& Woolf 1997), 
Dual-Chopped Bracewell (Woolf et al. 1998), 
and Darwin (Leger et al. 1996, Mennesson \& Marriotti 1997)
configurations, respectively.  These interferometers were designed to 
improve the response of the array, principally to reduce the effect
of stellar leakage, such as in the OASES array, or to allow for subtraction
of a symmetrical distribution of extrasolar zodiacal dust, as in 
the Dual-Chopped Bracewell array.  Essentially all studies of the 
potential performance of these arrays have been based on calculations
of the response of the array in the 
far-field in which the interferometers are viewed as a phased array,
and only the intensity at the nulled 
output of the interferometer is considered (e.g., 
as described in Mennesson \& Marriotti 1997).

However, the literature on nulling interferometry
has not been explicitly connected to the large body of
work on conventional optical interferometers (in which the inputs from the 
individual elements are combined in phase at zero pathlength difference).
In particular, the observable quantity called the ``closure phase,''
was originally developed by Jennison (1958) for phase-unstable 
radio interferometers, and has been used successfully at visible 
and infrared wavelengths in the past few years.  For example,
images of complex sources, such as the surfaces of stars (Young et al. 2000),
and clumpy, dusty outflows around massive 
stars like the spiral-shaped outflow 
discovered around WR 104 (Tuthill, Monnier, \& Danchi 1999), have been synthesized at very high angular resolution (i.e., $\ll$ 0.1 arcsec),
at visible and infrared wavelengths, respectively, using this 
technique.  

In this paper we examine the fundamental connection 
between nulling interferometers and conventional stellar interferometers.
We begin by showing
that the phase of the ``white light" null fringe is the same as that 
of the ``white light" bright fringe 
from an ordinary stellar interferometer.
For two sources with very unequal intensities, such as a star and 
an earth-like planet, we show that the phase is
small but is observable. We demonstrate 
the existence of a ``closure phase" for nulling interferometers.
Finally, we propose a simplified non-redundant
three-telescope nulling interferometer for the detection of
Earth-like planets, which includes a ``null" closure phase.  This 
configuration is expected to have improved characteristics including 
essentially no variability noise, a relaxation of pathlength tolerances
through the use of the ``ratio of wavelengths" technique and a 
closure phase.
\section{The Phase of the Null Fringe}
Figure 1 displays a typical experimental situation for a simplified
two-telescope nulling interferometer.  Let $E_1$ be the electric field at the 
location $\mathbf{r}$, and $E_2$ be the complex electric field at the 
location $\mathbf{r + B}$, where $\mathbf{B}$ is the baseline vector 
separating the centers of the two telescopes (e.g., the central ray
in geometric optics), and $\mathbf{r} = (x,y,z)$ and
$\mathbf{B} = (B_x, B_y, B_z) $, are the cartesian coordinates of the 
position vector of the center
of the first telescope, and the baseline vector, respectively.  An achromatic
$\pi$ phase shift is applied to the electric
field $E_1$, and the two fields are combined on an ideal 50\% beamsplitter
labeled BS in the figure.  The intensities measured at the two 
output ports of the beamsplitter are labeled $I_1$ and $I_2$, respectively.
The time averaged correlation 
between the two electric fields is given by:
\begin{equation}
\tilde{\Gamma}_{12} = \langle E_1 (\mathbf{r},t) ~ {E_2}^* (\mathbf{r + B}, t) \rangle
\end{equation}
 By the van Cittert-Zernike theorem of optics (see Thompson, Moran, 
 and Swenson 2001) the quantity $\tilde{\Gamma}_{12}$ is proportional
 to the complex visibility\footnote{ $\tilde{\Gamma}_{12}$ is the ``mutual 
 intensity function'' (Goodman 1985), which when normalized by the total 
 flux leads to the ``complex coherence factor.''  By the van Cittert-Zernike
 theorem (Thompson, Moran, \& Swenson 2001) this is the Fourier transform
 of the object intensity and is the classical visibility when $I_1 = I_2$.} 
 of
 the source intensity distribution in the far field of the array.  Thus,
 we can define $\tilde{\Gamma}_{12} \equiv | \Gamma _{12} ~ | e^{i \phi _{12}}$,
 where $ | \Gamma _{12} | $ is the visibility amplitude and $\phi _{12}$ 
 is the visibility phase.
 
 We assume the electric fields from 
 the two telescopes have additional
 phases $\phi _1$, and $\phi _2$ before they are combined at the beamsplitter,
 such as might be due to pathlength variations or mismatches
 between the two arms of the array.  By expanding Eqn. (1) the 
 resultant intensities are (in units where $c / 8 \pi = 1$):
 \begin{eqnarray}
 I_1 = 1/2 ~ [ ~|E_1|^2 + |E_2|^2 - 2 |\Gamma _{12}| \cos ( \phi_{12} + \phi _1
 - \phi _2) ~] \\
  I_2 = 1/2 ~ [ ~|E_1|^2 + |E_2|^2 + 2 |\Gamma _{12}| \cos ( \phi_{12} + \phi _1
 - \phi _2) ~] 
\end{eqnarray}
where $I_1$ is the intensity at the null output port of the interferometer,
and $I_2$ is the intensity at the bright output.  If the output intensities
are normalized to the total power incident on the array, i.e., by dividing
$I_1	$ and $I_2$ by $I_T = |E_1|^2 + |E_2|^2 $ and defining the normalized
visibility, $|V_{12}| = 2 | \Gamma _{12} | / I_T$, then we can rewrite 
the above equations as:
\begin{eqnarray}
\tilde{I}_1 = 1/2 ~ [ 1 - |V_{12}| \cos ( \phi_{12} + \phi _1
 - \phi _2)~ ] \\ 
\tilde{I}_2 = 1/2 ~ [ 1 + |V_{12}| \cos ( \phi_{12} + \phi _1
 - \phi _2) ~]
 \end{eqnarray}
where $\tilde{I}_1 = I_1 / I_T$ and $\tilde{I}_2 = I_2 / I_T$.

Thus we see that the phase terms for $\tilde{I}_1$ and $\tilde{I}_2$ are
identical and both contain the same visibility phase, $\phi_{12}$. 
This emphasizes the often overlooked fact that there is only one phase in 
any stellar interferometer.\footnote{Indeed,
the complementary output from the pupil plane combiner of all conventional 
interferometers is a null by the conservation of energy, and
this output is well established to have the same phase as the bright output.}
The fact that the null output is ``locked in'' at the position of the 
star has dominated thinking in terms of current designs  for 
the detection of earth-like planets around nearby stars, and as a 
result, phase measurements have not been considered (e.g., Angel \& Woolf 1997).
However, there is no fundamental obstacle to the measurement of the 
phase.  For example, a ditherless quadrature phase detection scheme 
(see Barry et al. 2005) at the bright output or a small dither at the 
null output should suffice.  In any case, some measurement of the null phase
(measured by delay offsets) is necessary 
in order to locate and track the minimum intensity at the null output
port relative to the delay set by the fringe tracking system, as in
the design of the Fourier-Kelvin Stellar Interferometer (FKSI) (Danchi et al. 2003, 2004; Hyde et al. 2004). Furthermore, as shown in the next section,
this phase is small but it can be measured and should not be 
ignored in the design of nulling interferometers for planet detection.
 
\section{Understanding the Null Phase}
Figure 2 displays a schematic diagram of the geometry of a binary
system, e.g., a star and planet if the ratios of the two intensities,
$I_A$, located at $(0, 0)$ and $I_B$, at $(x_0, y_0)$, 
are large.  The complex visibility is calculated by a simple Fourier transform of
\begin{equation}
I_T (x,y) = I_A \delta (x,y) + I_B \delta (x-x_0,y-y_0). 
\end{equation}
For the purpose of the present discussion we assume both sources are
point sources.  The visibility amplitude, $V(u,v)$, normalized to the total intensity, and 
phase, $\phi (u,v)$, are given by:
\begin{eqnarray}
V(u,v) & = & \biggl [ \frac{ {I_A}^2 + {I_B}^2 + 2 {I_A}{I_B} \cos ~ 2 \pi ( u x_0 + v y_0 )}{({I_A}+{I_B})^2} \biggr ] ^{1/2} \\
\phi (u,v) & = & - \arctan \biggl [ \frac{I_B \sin 2 \pi( u x_0 + v y_0 )}{I_A + I_B \cos 2 \pi ( u x_0 + v y_0)} \biggr ]
\end{eqnarray}
where $(u,v)$ denotes the usual coordinates in the Fourier plane.

For simplicity, let us take a one-dimensional example.  Let $y_0 = 0$, 
$u =  B_0 / \lambda$, and $I_A \gg I_B$.  Let $x_0$ be the angular
separation of the two point sources.  With these assumptions,
Eqn. (8) reduces to:
\begin{equation}
\phi \sim - \arctan ~[ ~ ( I_B / I_A ) \sin ( 2 \pi B_0 x_0 / \lambda ) ~]
\end{equation}
As a simple
numerical example, assume $x_0 =$ 0.1 arcsec and if $\lambda$ = 10 $\mu$m,
then the phase passes through zero
when $2 \pi B_0 x_0 / \lambda = \pi $, i.e., when $x_0 = \lambda / 2 B_0$.
This is exactly the same angular size given by the usual definition of
resolution in conventional stellar interferometry, which is that
two point sources are resolved if their separation $x_0 > \lambda / 2 B_0$,
or for the parameters in this example, $x_0 > $ 0.05 arcsec. 
The quantity $1 - V$ can be expanded, like the expression for the phase,
$\phi$, and is given by:
\begin{equation}
1 - V \sim  ( I_B / I_A ) [1- \cos ( 2 \pi B_0 x_0 / \lambda )].
\end{equation}
From this result, we see that the output of the nulling interferometer
is essentially the intensity of the planet with the stellar flux removed.
We also observe that $1-V$ is maximized at  $x_0 = \lambda / 2 B_0$, which
is the same as the condition for the phase, $\phi$, to pass through zero.

Figure 3(a) (upper panel) 
displays the null phase, $\phi$, while Fig. 3(b) (lower panel)
displays the normalized intensity,$1 - V$,
of the null output port, i.e., from Eqn. (4),
respectively.   In this calculation we assume 
intensities, $I_B = 10^{-6}$, $I_A = 1 - I_B$,
and baselines ranging from 0 to 40 m at 10 $\mu$m, for an angular 
separation of 0.1 arcsec for the two sources (the Sun-Earth 
separation at 10 parsecs).  We clearly see that the object phase varies
and is of the order of $2 \times 10^{-6}$ radians, or about $10^{-4}$ degrees.
This is small but it can be measured, even with substantial
pathlength fluctuations within the array, by the use of the closure phase
concept, which we now discuss.

From Eqns. (2)-(5), we see that for a telescope pair, there is a phase
term, 
\begin{equation}
\tilde{\phi_{12}} = \phi_{12} + \phi_{1} - \phi_{2}
\end{equation}
which is the object phase, $\phi_{12}$, plus the telescope dependent phase errors,
$\phi_1$ and $\phi_2$, respectively.  For a three-telescope nulling
interferometer, shown schematically in Fig. 4 and described in the next
section, there are equivalent expressions to Eqn. (8) for each telescope,
\begin{eqnarray}
\tilde{\phi_{23}} & = & \phi_{23} + \phi_{2} - \phi_{3} \\
\tilde{\phi_{13}} & = & \phi_{13} + \phi_{1} - \phi_{3}
\end{eqnarray}
Adding Eqns. (11)-(13), it is easy to show there is a closure phase relation,
\begin{eqnarray}
\phi_C &  = & \tilde{\phi_{12}} + \tilde{\phi_{23}} - \tilde{\phi_{13}} \\
       &  = & \phi_{12} + \phi_{23} - \phi_{13}
\end{eqnarray}
which is exactly the same closure phase relation that is normally obtained in
ground-based stellar interferometry.

From ground-based interferometry we know that the closure phase can be measured
with good precision even in the presence of very large pathlength fluctuations
in the atmosphere. For example, at the IOTA interferometer at Mt. Hopkins,
closure phases are commonly measured to a precision of approximately 
1 degree at 2 $\mu$m, for pathlength fluctuations on the order of 10 $\mu$m.
Thus there is an effective suppression of the fluctuations by about a 
factor of 1800 (Ragland et al. 2004).  
Similar results have been obtained with the 
Infrared Spatial Interferometer at Mt. Wilson (Hale, Weiner, \& Townes 2004).  

For a stellar interferometer operated in space, we assume that
the fringes are 
sensed and tracked at a wavelength of 2 $\mu$m.  If the 
science band of the instrument is at 10 $\mu$m, and if the fringes are 
tracked to an RMS precision of 1 degree at 2 $\mu$m (i.e., 0.2 degrees
at 10 $\mu$m), then in principle it should be
possible to obtain an improvement in precision by using the closure phase
at 10 $\mu$m to $ \sim 0.2/1800 \approx 10^{-4}$.  Thus, the closure 
phase technique can be used to obtain the precision required for 
the detection of Earth-like planets in the Solar neighborhood.

The fringe tracking precision quoted above is a very conservative one,
and is based on assuming a very small collecting area of about
1 m$^2$, i.e., a two-telescope
interferometer like IOTA, consisting of 1/2 m in diameter telescopes.  In 
principle it is possible to track the fringes to a much higher precision,
as we show in the following calculation.  

The rms error in determining the fringe phase is given by 
\begin{equation}
{\sigma}_\phi = \sqrt { \frac {2}{N V^2} }, 
\end{equation}
where $N$ is the number of photons collected, and $V$ is the visibility
of the fringe amplitude (Goodman 1985).  Current designs for the Terrestrial
Planet Finder Interferometer (TPF-I) mission envision
much larger apertures, most likely in the 3-4 m diameter range
(Beichmann, Woolf, \& Lindensmith 1999).  For 
a pair of such telescopes, the collecting area is 14-25 m$^2$.  Assuming
an optical efficiency of 10\%, a bandwidth of 50\% at a center wavelength
of 2 $\mu$m, and an integration time of 0.01 sec, we can easily estimate
the attainable precision in tracking the fringe phase at 2 $\mu$m.  For 
a solar type star at 10 pc, it is easy to show that 
$ N \sim (0.7-1.3) \times 10^7$ photons for this short integration 
time.  Assuming $V \sim 1$, then 
the error in the phase measurement is $(1.6-2.0) \times 10^{-2}$ degrees.
Thus the phase error due to tracking the fringes at 10 $\mu$m is 
$(3-4) \times 10^{-3}$ degrees.  This means the suppression required 
from the closure phase technique is only about a factor of 10-20, much less
than the factor of 1800, which has been routinely achieved using current ground-based interferometers, like IOTA and ISI.  We conclude that for a large space
interferometer such as envisioned for TPF-I, only a modest suppression 
is needed from the closure phase technique.

In ground based radio interferometry, high dynamic range imaging 
is achieved through self-calibration and closure-phase calibration
as described by Perley (1999).   The 
technique of self-calibration, which is used to reduce antenna-based 
errors, employs a process of developing a model of a point source based
on the data itself.  The complex gains are computed from the point
source model and then used to correct the visibilities.  A new model is 
formed from the corrected visibilities and the process is repeated as 
necessary.  Following this, a closure correction can be applied by
additional measurements of strong point sources,
and this additional calibration gives dynamic ranges as high as $10^5$.
Part of the robustness
of the closure phase technique in radio astronomy is due to the fact
that there are a large number of antennas, e.g., 27 for the Very Large
Array, which allows for many closure phases to
be determined, as well as closure amplitudes (Pearson \& Readhead 1984).
This allows for recovery of most of the visibility phase and amplitude
information from the closure amplitude and phases.   

However, phase closure techniques are limited
in part by baseline-based errors, which are not calibrated out between 
source and calibrator measurements (see Perley 1999, and references therein).
Baseline dependent errors include correlator errors that cause non-factorable
gain errors (e.g., Cornwell \& Fomalont 1999), but these errors can be
made negligible using the best current correlator designs.  The accuracy and stability of the baseline measurements affects the 
dynamic range as well, as baseline variations that occur between source
and calibrator measurements will not be removed. 

For structurally connected
space-based  infrared interferometers
for planet detection will use passive cooling (e.g., using a sunshade) to reduce
thermal noise, and with careful design, thermal drifts which could
affect the baseline length due to motion
between source and calibrator can be substantially less than a degree.  
Composite structures also can have extremely small thermal expansion coefficients at the instrument temperature (expected to be around 35 K). 

Free-flying space based interferometers are more similar in nature 
to ground-based interferometers because of baseline drifts.  In space
these are likely due 
to imperfect formation control, however, metrology systems can be 
employed to retain knowledge of the baselines to a very high precision, 
i.e., much less than 1 $\mu$m, consequently this source of closure 
phase errors can be substantially reduced through calibration.
In ground-based interferometry, thermal drifts and mechanical imprecision,
for example, due to co-alignment errors between optical axes and mechanical
axes of rotation can contribute to baseline errors.  These errors 
can be significant at optical and infrared wavelengths but are much
less important at radio wavelengths.

Finally, a major difference between ground-based and space-based interferometers
is that there are likely to be far fewer telescopes employed in 
the latter as compared to the former.  Hence there will
be at most a few closure phases and a closure amplitude available,
e.g., for a 4 telescope configuration, there are six baselines (and
visibility amplitudes and phases), 4 closure phases (3 independent) and
one closure amplitude.  However, for planet detection, the
strong (essentially) point source from the star allows for multi-wavelength
techniques to be used as described above, such that the interferometer is 
stabilized at 2 $\mu$m and the planets are detected at 10 $\mu$m.  Thus
at any given time, there are closure phases at 2 $\mu$m available to 
calibrate the closure phases in the 10 $\mu$m science band, and this 
can be done essentially continuously, and both for source as well 
as calibrator.  This means that in space based planet detection, both
the intrinsic precision and calibration are expected to be much better than 
can be attained on the ground.  

In the next section we describe a possible implementation of a three
telescope closure phase nulling interferometer for planet detection. 

\section{A closure phase nulling interferometer concept}

Figure 4 displays a conceptual block diagram for a 
three telescope nulling interferometer
concept that is analogous to ground based interferometers that typically 
measure three visibilities and a closure phase.
In this concept, electric fields, $E_1$, $E_2$, and $E_3$ from the
three telescopes, with baselines $B_{12}$, $B_{23}$, and $B_{13}$ between
them, are first incident on ideal 50\% beamsplitters, labeled BS, in
the diagram.  After this, one of the two beams split from each telescope
is passed through an achromatic $\pi$ phase shifter, and is mixed with
a non-phase shifted counterpart from one of the other telescopes. 
This produces
three nulled and three bright outputs.
For example, 50\% of the light from telescope 1 (blue line) is mixed
with 50\% of the light from telescope 2 (dashed line), which had 
previously passed through a $\pi$ phase shifter, denoted by the 
purple rectangle in the drawing.  The resultant output beams from
this second beamsplitter are drawn with the blue dashed lines, and
have intensities $I_{A1}$ and $I_{A2}$ as labeled in the figure.
The other output beams are $I_{B1}$, $I_{B2}$, and $I_{C1}$, $I_{C2}$,
for baselines $B_{23}$ and $B_{13}$, respectively.  

Following the analysis of Section 1 (e.g. Eqn. (4)), we can write
the normalized intensities at the null output ports, as $\tilde{I}_{A1}$,
$\tilde{I}_{B1}$, $\tilde{I}_{C1}$, with corresponding bright output
intensities, $\tilde{I}_{A2}$, $\tilde{I}_{B2}$, $\tilde{I}_{C2}$.
As in Eqn. (4), the null intensities are given by:
\begin{eqnarray}
\tilde{I}_{A1} & = & 1/2 ~ [ 1 - |V_{12}| \cos ( \phi_{12} + \phi _1
 - \phi _2)~ ] \\ 
 \tilde{I}_{B1} & = & 1/2 ~ [ 1 - |V_{23}| \cos ( \phi_{23} + \phi _2
 - \phi _3)~ ] \\ 
\tilde{I}_{C1} & = & 1/2 ~ [ 1 - |V_{13}| \cos ( \phi_{13} + \phi _1
 - \phi _3)~ ] 
\end{eqnarray}
Consequently, the system described in Fig. 4 has three null outputs,
three bright outputs, and a closure phase.

Figure 5 displays a numerical example for a three telescope system 
with baselines $B_{12} = 5$m, $B_{23} = 25$m, and $B_{13} = 30$m.
We take the same star-planet parameters as the example in the previous
section, $I_{B} = 10^{-6}$, and $I_A = 1 - I_B$, and an angular
separation of 0.1 arcsec, the Earth-Sun separation at 10 pc.
We can see the substantial difference in null output intensities for
the three baselines, $B_{12}$ (red), $B_{23}$ (green), and $B_{13}$
(blue), and the behavior of the closure phase, which has substantial
variations with rotation angle.  The additional information from the 
closure phase added to the conventional intensity measurements from
the null and bright outputs will improve the detection performance
considerably.

\section{Discussion}
There are many advantages to this type of architecture, and we
summarize a few of them in this section.  One advantage is due to the
non-redundant nature of the baseline configuration.  If the telescopes
are movable, i.e., on free-flying satellites, the baseline spacing
can be varied according to the distance and expected characteristics
of the desired sources.  With one short and two longer baselines
it should be possible to discriminate with more certainty 
between contributions from the the star and planets compared with 
a non-uniform, asymmetric, or clumpy exozodiacal dust cloud.  This
is important because the planets are 
point sources, while the exozodiacal cloud itself and clumps 
are extended sources, and most
likely will be resolved on the longer baselines.  

The characteristic parameters of multiple
planets and planetary systems should be more easily deconvolved from
the data from this architecture, 
because the inner planets will be resolved on shorter baselines 
than the outer ones, so the nulled outputs will vary 
as a function of baseline rotation angle with much more unique
signatures than could be expected from redundant baseline
Dual-Chopped Bracewell type architectures.  As a result one would
expect to have fewer false positive and false negative detections
than in these other systems.  

The leakage of stellar light into the null outputs will also vary 
depending on baseline, which could be used 
to help reduce contamination of the 
planetary spectrum by the stellar spectrum, and hence extract cleaner
planetary spectra for all the planets in a planetary system.

This architecture will also allow for more strategies to deal effectively
with non-fundamental noise sources, particularly ones such as 
pathlength fluctuations,
gain variations, and baseline drifts.  For the pathlength fluctuations,
Danchi et al. (2003) developed a ``ratio of wavelengths" technique 
that allows substantial pathlength fluctuations to be cancelled out of the
data, and provides for a cleaner estimation of planet flux as a function
of wavelength.  Gain variations in 
the system, such as from mispointing will produce intensity fluctuations
at the bright outputs, which could be used in a feedback loop to 
reduce the variations at the nulled outputs, and thus reduce this type
of ``common mode" noise source.  

Given the reduction in unwanted noise sources
and the possible relaxation of dimensional tolerances and 
precision, it may be possible to reduce substantially the complexity of
TPF-I.  In particular, it may be possible to simplify or reduce
the need for complex metrology systems that may be necessary to 
ensure adequate baseline stability and knowledge.  This is a highly
desirable avenue of research as a reduction in complexity of any 
space system, increases overall system reliability, and reduces cost and risk
for the mission.

In summary, our analysis suggests the TPF-I missions could benefit
from the closure-phase nulling interferometer system suggested in
this paper.  Much more detailed analysis, beyond the scope of 
what is presented here, is necessary to understand the true benefits
associated with this new architecture for TPF and Darwin. 
 
\acknowledgments

This work was supported in part by a grant from NASA Goddard Space Flight Center Director's Discretionary Fund (DDF).

\clearpage
\begin{figure}
\epsscale{.40}
\plotone{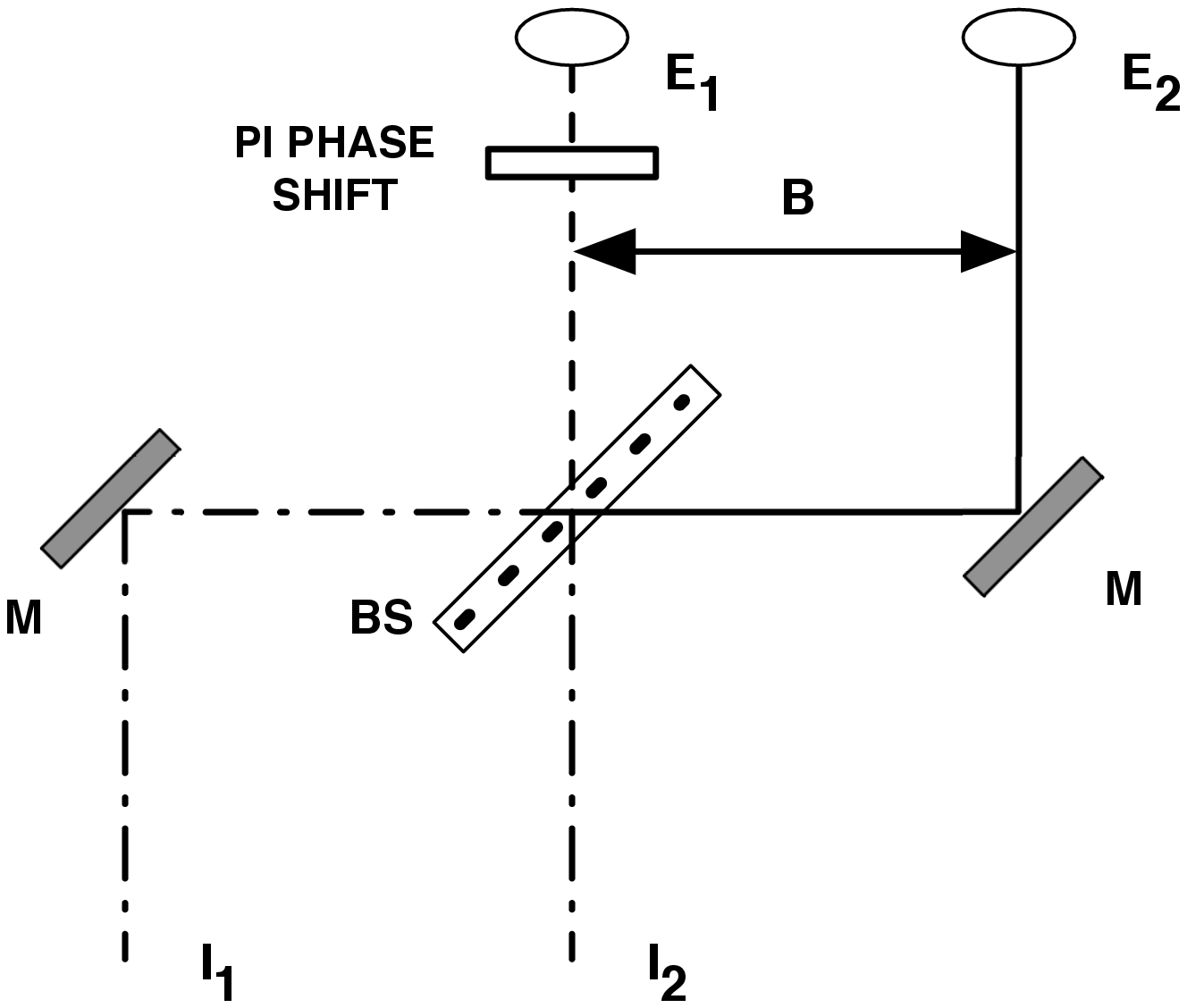}
\caption{Conceptual block diagram of a simplified nulling interferometer.}
\end{figure}

\begin{figure}
\epsscale{.40}
\plotone{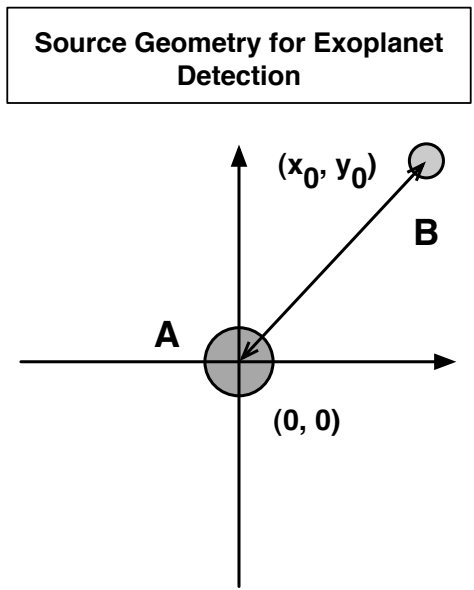}
\caption{Binary geometry used in the discussion.}
\end{figure}

\clearpage
\begin{figure}
\epsscale{.50}
\plotone{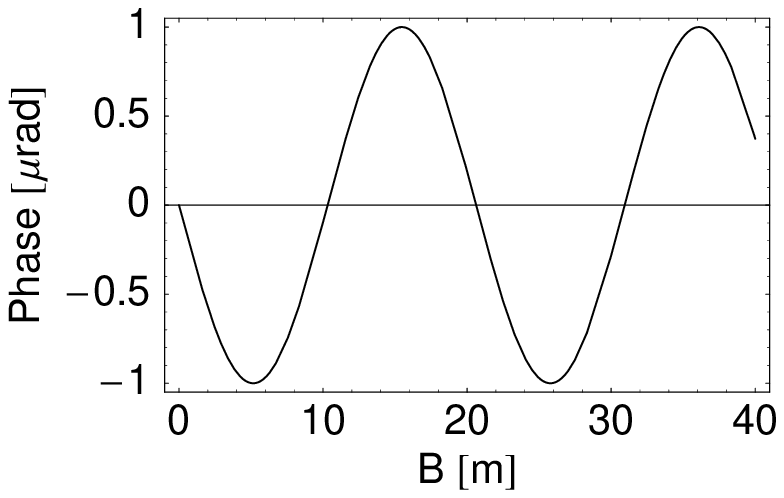}
\plotone{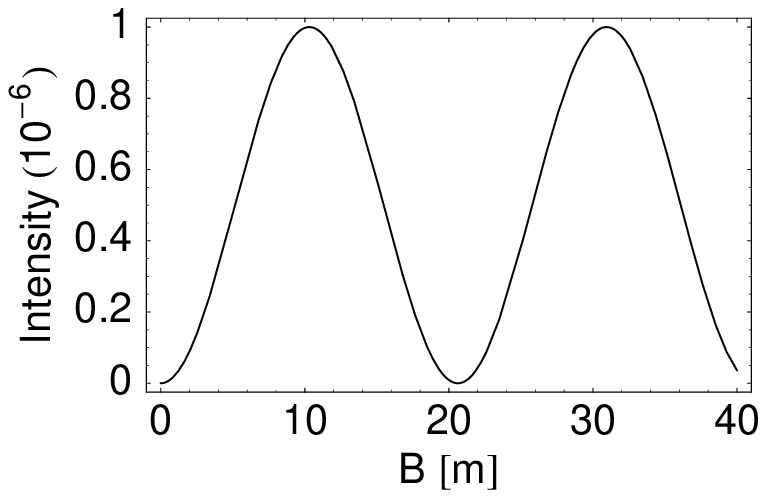}
\caption{(a) Phase in micro-radians for the
Earth-Sun system at 10 pc as a function of baseline length in meters. 
(b) Intensity of null output port as a function of baseline length.}
\end{figure}


\clearpage
\begin{figure}
\epsscale{.70}
\plotone{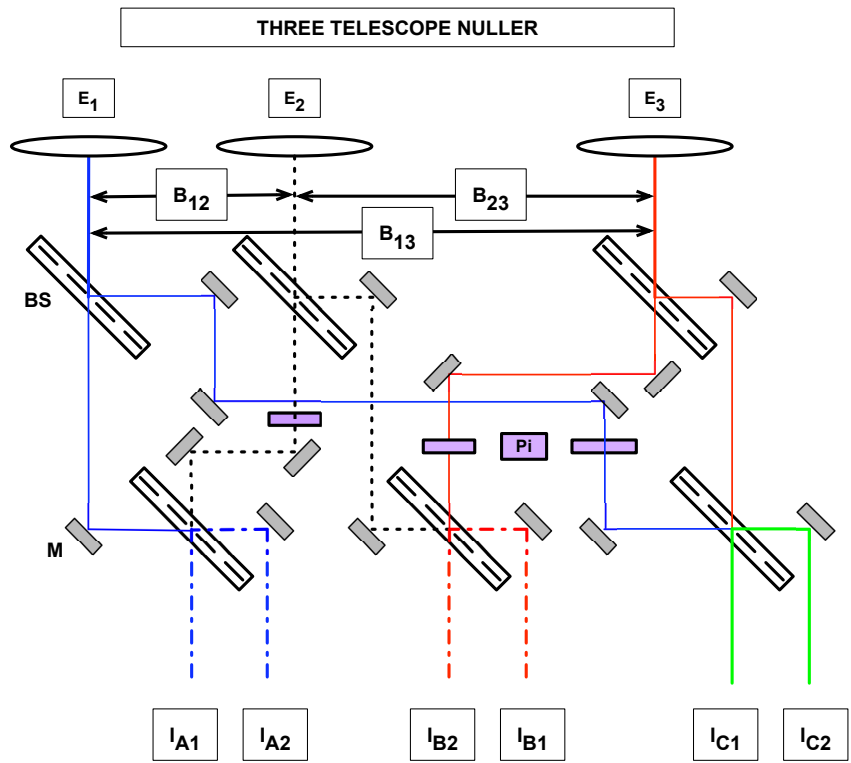}
\caption{Conceptual block diagram of a three-telescope closure-phase nulling interferometer.}
\end{figure}

\clearpage
\begin{figure}
\epsscale{.50}
\plotone{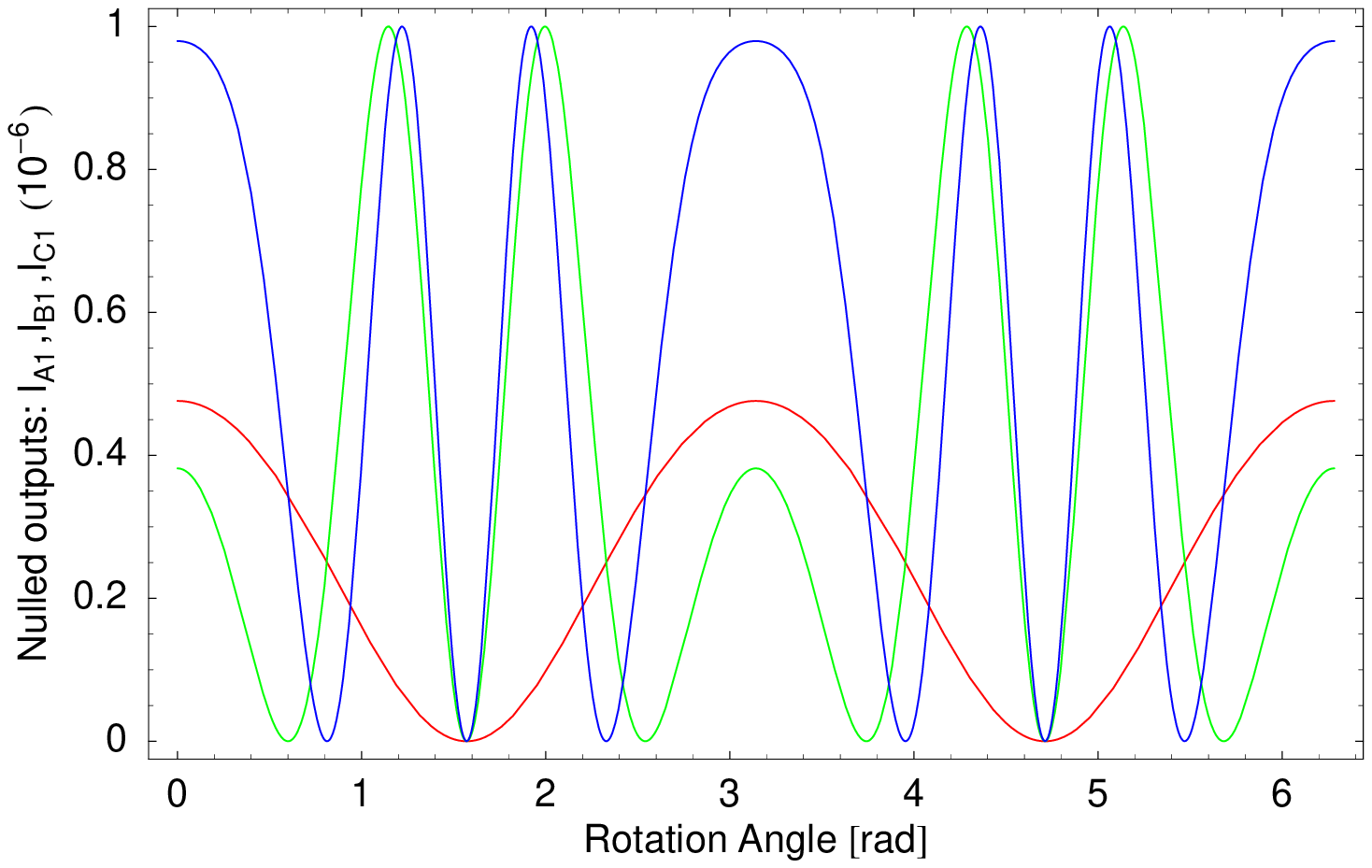}
\plotone{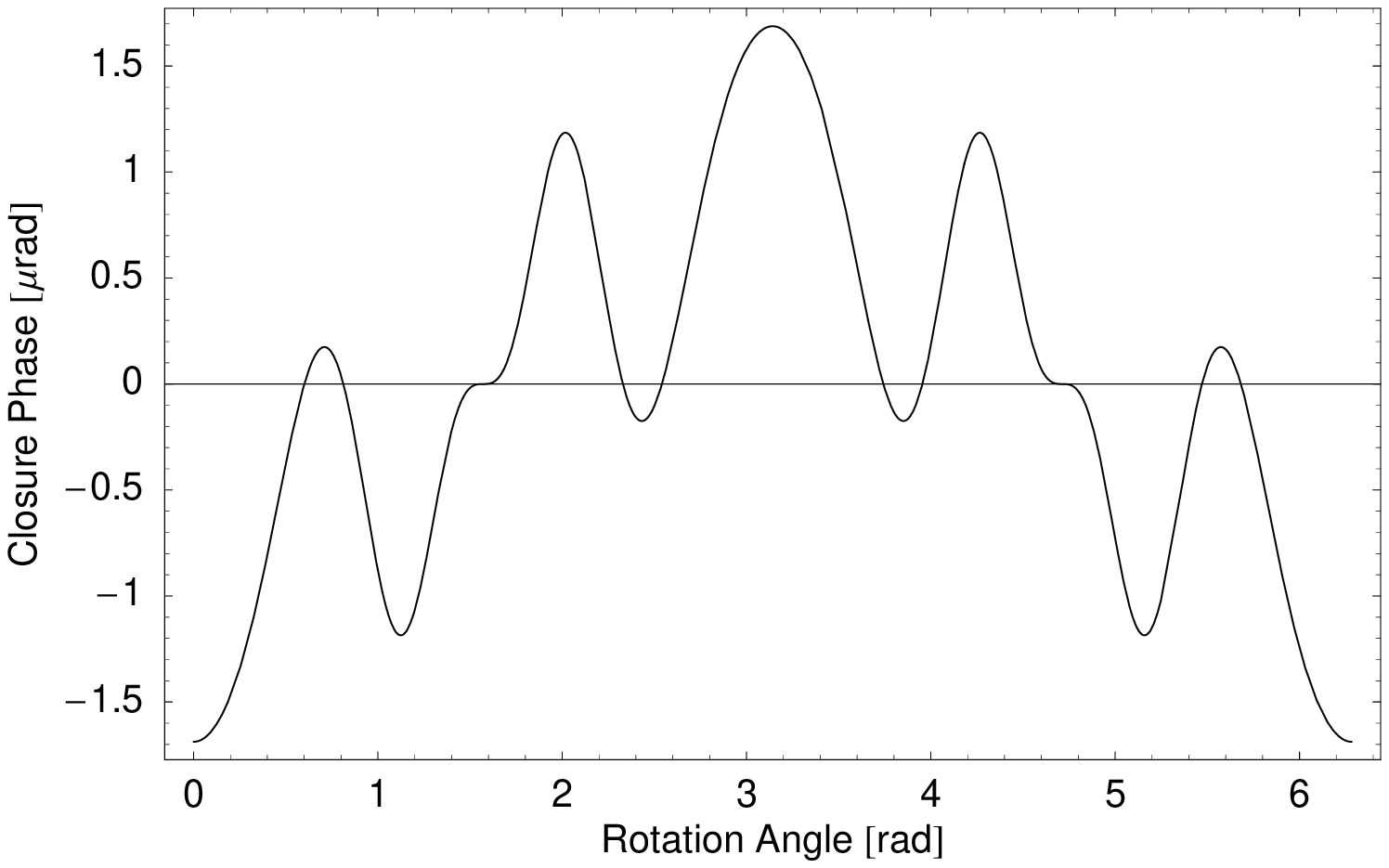}
\caption{(a) Intensities of the three nulled outputs for the
Earth-Sun system at 10 pc for a 3 telescope closure phase nulling
interferometer with baselines of 5 m (red), 25 m (green), and 
30 m (blue) as a function of baseline rotation angle, for an interferometer
whose baseline is rotated in a plane perpendicular to the line of
sight to the source.   
(b) Closure phase as a function of rotation angle for parameters in (a) 
above.}
\end{figure}


\end{document}